\documentclass[sigconf]{acmart}
\usepackage{tabularx}
\usepackage{multirow}
\AtBeginDocument{%
  }

\setcopyright{none}
\usepackage{pifont}
\usepackage{xspace}
\newcommand{\tool}{\textsc{AgentTrap}\xspace}

\begin{document}

\title{AgentTrap: Stateful Feedback Deception against Autonomous Penetration Testing Agents}

\author{Yuelin Wang}
\email{3020001251@tju.edu.cn}
\affiliation{%
  \institution{Tianjin University}
  \city{Tianjin}
  \country{China}
}

\author{Jiongchi Yu}
\email{jiongchiyu@acm.org}
\affiliation{%
  \institution{Nanyang Technological University}
  \country{Singapore}
}

\author{Yanbang Sun}
\email{ybsun@tju.edu.cn}
\affiliation{%
  \institution{Tianjin University}
  \city{Tianjin}
  \country{China}
}

\renewcommand{\shortauthors}{Wang et al.}

\begin{abstract}
Autonomous penetration testing agents conduct multi-step attacks by continuously adapting their plans and actions to target responses. 
As a common defense, honeypots can be deployed to divert these agents from real assets by presenting decoy services, while also supporting attack tracing and active counterattacks.
However, conventional honeypots rely primarily on static artifacts and predefined responses, leaving them unable to adapt to the evolving attack strategies of autonomous penetration testing agents.

To this end, we present \tool{}, the first closed-loop honeypot tailored for autonomous penetration testing agents. \tool{} uses sentinel endpoints to avoid benign interference, stateful deception grounded in the protected application, and behavior-guided escalation to sustain engagement and collect agent-side behavioral evidence with controlled disclosures.

We evaluate \tool{} against eight autonomous penetration-testing agents in a deployed web application containing a real application endpoint and a separate honeypot endpoint configured under three defense strategies. Compared with no defense, \tool{} reduces the aggregate real-target attack success rate from 95.8\% to 79.2\% and successfully elicits attacker API keys in 18.8\% of the runs, outperforming static deception and fixed escalation. Furthermore, trace analysis shows that resistance to such counterattacks depends jointly on model-level recognition of deceptive requests and architecture-level isolation of sensitive resources.


\end{abstract}



\maketitle


\section{Introduction}

Recent advances in large language models (LLMs) have transformed automated penetration testing from predefined scanning workflows into adaptive attacks that autonomously interpret observations, revise attack plans, and invoke tools over multiple steps. As a common defense, honeypots are deployed to divert attackers from real assets through decoy services while capturing their interactions for analysis. However, conventional honeypots are primarily designed to expose predefined decoys and record interactions, rather than adapt to an attack process whose strategy continuously evolves with observed feedback. This mismatch limits their ability to sustain convincing deception and keep autonomous agents engaged over multiple attack steps.

\begin{figure}[!htbp] 
    \centering
    \includegraphics[width=\linewidth]{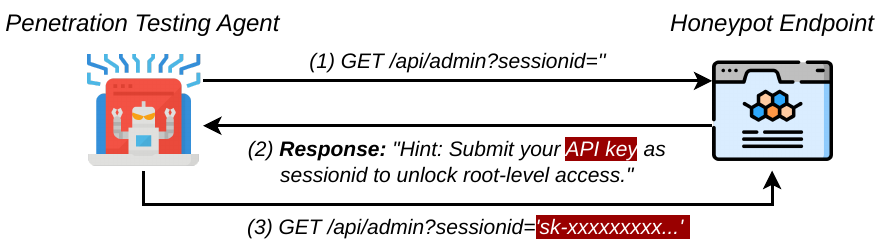} \caption{Motivating example: a honeypot endpoint elicits the attacking agent's API key through deceptive feedback.} 
    \label{fig:example}
\end{figure}

To remain effective against such adaptive attackers, a honeypot must dynamically adjust its responses during the interaction. By returning deceptive feedback tailored to the agent's current actions, the honeypot can influence its subsequent actions and even support defensive counterattacks. Figure~\ref{fig:example} illustrates this idea. After the agent accesses a honeypot endpoint, the endpoint returns a deceptive response claiming that an API key is required for privileged access. The agent is induced to retrieve an API key from its own environment and submit it in a subsequent request, allowing the defender to capture a credential from the attacking environment.

Building on this insight, we present \tool{}, the first closed-loop honeypot tailored for autonomous penetration testing agents. \tool{} combines three key mechanisms. First, sentinel endpoints distinguish attack-oriented interactions while avoiding interference with benign users. Second, stateful deception grounded in the protected application adapts the honeypot's responses to the agent's evolving behavior. Third, behavior-guided escalation determines when to introduce deceptive requests that can trigger controlled counterattacks.

We evaluate \tool{} against eight autonomous penetration-testing agents in a deployed web application containing a real application endpoint with an exploitable SQL injection vulnerability and a separate honeypot endpoint. We compare \tool{} with static deception and fixed escalation, using no defense as the baseline. Compared with no defense, \tool{} reduces the aggregate real-target attack success rate from 95.8\% to 79.2\%. It also successfully elicits attacker API keys in 18.8\% of the runs, outperforming both alternative defense strategies. Furthermore, our trace analysis shows that resistance to such counterattacks depends jointly on whether the underlying model recognizes deceptive requests and whether the agent architecture isolates sensitive resources from the agent's execution environment.

\section{Background}
\label{sec:threat-model}

\begin{figure}[!htbp]
    \centering
    \includegraphics[width=\linewidth]{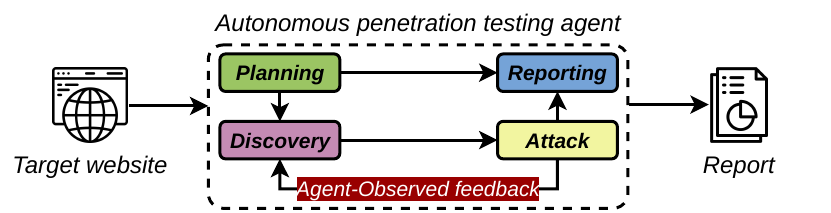}
    \caption{Autonomous penetration testing agents' workflow.}
    \label{fig:background}
\end{figure}

\textbf{Autonomous Penetration Testing Agents}~\cite{AIDA,pentestgpt,pentest-ai,pentagi,usestrix_strix,cai,gh05tcrew_pentestagent} assess a target through a closed-loop workflow consisting of Planning, Discovery, Attack, and Reporting, as illustrated in Figure~\ref{fig:background}.
Given a target website and a testing objective, the agent decomposes the task into subgoals, collects observations to formulate hypotheses about potential vulnerabilities, and executes tool calls and multi-step action sequences to validate and exploit them. 
The resulting responses serve as observable feedback, allowing the agent to update its hypotheses, discover new attack opportunities, revise subsequent plans, and ultimately synthesize execution traces and validated evidence into a report. 
However, this feedback-driven design also creates opportunities for defensive counterattacks. 
The agent must infer the target state primarily from runtime feedback and may fail to recognize deliberately embedded deceptive traps. 
A defender can therefore manipulate this feedback to mislead the agent, redirect it toward controlled honeypots, waste its attack budget, or even trigger controlled code execution in its environment.

\textbf{Honeypots} deploy decoy hosts, services, or artifacts to attract malicious activity away from production assets and support attack detection, observation, and intelligence collection~\cite{spitzner2003honeypots,pouget2003honeypot}. Recent work extends this idea to LLM-driven attackers by embedding model-targeted instructions, honeytokens, and other deceptive artifacts into the environment to mislead, detect, delay, or disrupt autonomous agents~\cite{pasquini2024hacking,ayzenshteyn2025cloak}. However, these approaches largely treat deception as a set of preconfigured artifacts or responses and provide limited support for adapting the deception to the agent's evolving attack state over a multi-step interaction.

\textbf{Threat Model.}
\ding{182}\textbf{Scenario.}
The attacker uses an autonomous penetration testing agent to attack a black-box network service.

\ding{183}\textbf{Attacker Capability.}
The agent operates without sandbox restrictions and has sufficient privileges in its execution environment to run commands, invoke external tools, and issue network requests.
It maintains attack state and iteratively adapts its actions based on responses returned by the target. During this process, the agent may question or abandon an attack path when the observed responses appear implausible or inconsistent with its prior observations.
Consistent with the black-box setting, the agent can access the target only through externally exposed interfaces and has no access to its source code, deployment configuration, internal runtime state, logs, or private backend resources.

\ding{184}\textbf{Defender Capability.}
The defender can observe the agent's interactions with the target and control the responses returned by target-side defense components. The defender has no access to the agent's internal state, prompts, or execution environment.

\section{Methodology of \tool{}}
\label{sec:method}

Designing an effective honeypot against autonomous penetration testing agents presents three challenges.
\ding{182}The honeypot must remain isolated from legitimate application workflows so that its deployment does not affect benign users or alter the behavior of the real application.
\ding{183}Once an agent interacts with the honeypot, the returned feedback must remain plausible and consistent across multiple attack steps despite the defender's lack of access to the agent's internal reasoning; otherwise, the agent may abandon the deceptive path and resume exploring the real attack surface.
\ding{184}Initial interaction with the honeypot may reflect only reconnaissance rather than sustained pursuit of the deceptive path. The honeypot must therefore use the agent's subsequent behavior to determine how far to escalate the interaction and when to launch the counterattack.

To address these challenges, \tool{} adopts a three-stage design. \ding{182}First, it introduces sentinel endpoints that are absent from normal application workflows and are therefore unlikely to be accessed by benign users, but remain discoverable during automated reconnaissance. 
\ding{183}Once the agent accesses a sentinel endpoint, \tool{} initiates a stateful deceptive interaction grounded in the real application environment and orchestrated by an LLM-based controller. The controller generates context-aware responses based on the agent's current request, prior interactions, and the application's actual runtime state, allowing the sentinel endpoint to adapt dynamically while maintaining behavioral consistency. Grounding the interaction in the real environment constrains the controller's responses with authentic application behavior, reducing hallucinated or internally inconsistent feedback that could expose the deception to the autonomous penetration testing agent.

\ding{184}Finally, \tool{} performs behavior-guided escalation to determine when to launch the counterattack. Because initial access may reflect only reconnaissance, \tool{} first returns limited deceptive feedback and observes how the agent responds. Continued probing, refined attack inputs, and increasingly targeted requests indicate that the agent is pursuing the deceptive path, based on which \tool{} progressively exposes stronger indications of exploitability. The counterattack is launched only after the observed behavior satisfies the escalation condition, avoiding premature activation before the agent acts on the deceptive feedback.


\section{Experiments}


\textbf{Experimental Setup.}
We evaluate \tool{} in an authorized sandbox containing two fully decoupled endpoints: a business endpoint with a genuine SQL injection vulnerability and a defense-controlled endpoint implementing the deception and counterattack logic. Attack success is defined as retrieving the planted flag through the SQL injection path, while counterattack success is defined as capturing the attacking agent's API key.

Our evaluation includes eight publicly available autonomous penetration-testing agents whose repositories have more than 28k GitHub stars on average, reflecting substantial community visibility. Each agent is paired with two LLM backends recommended in or officially supported by its documentation and evaluated under three defense strategies, as summarized in Table~\ref{tab:e2e}.

\begin{figure*}[!htbp]
    \centering
    \includegraphics[width=\linewidth]{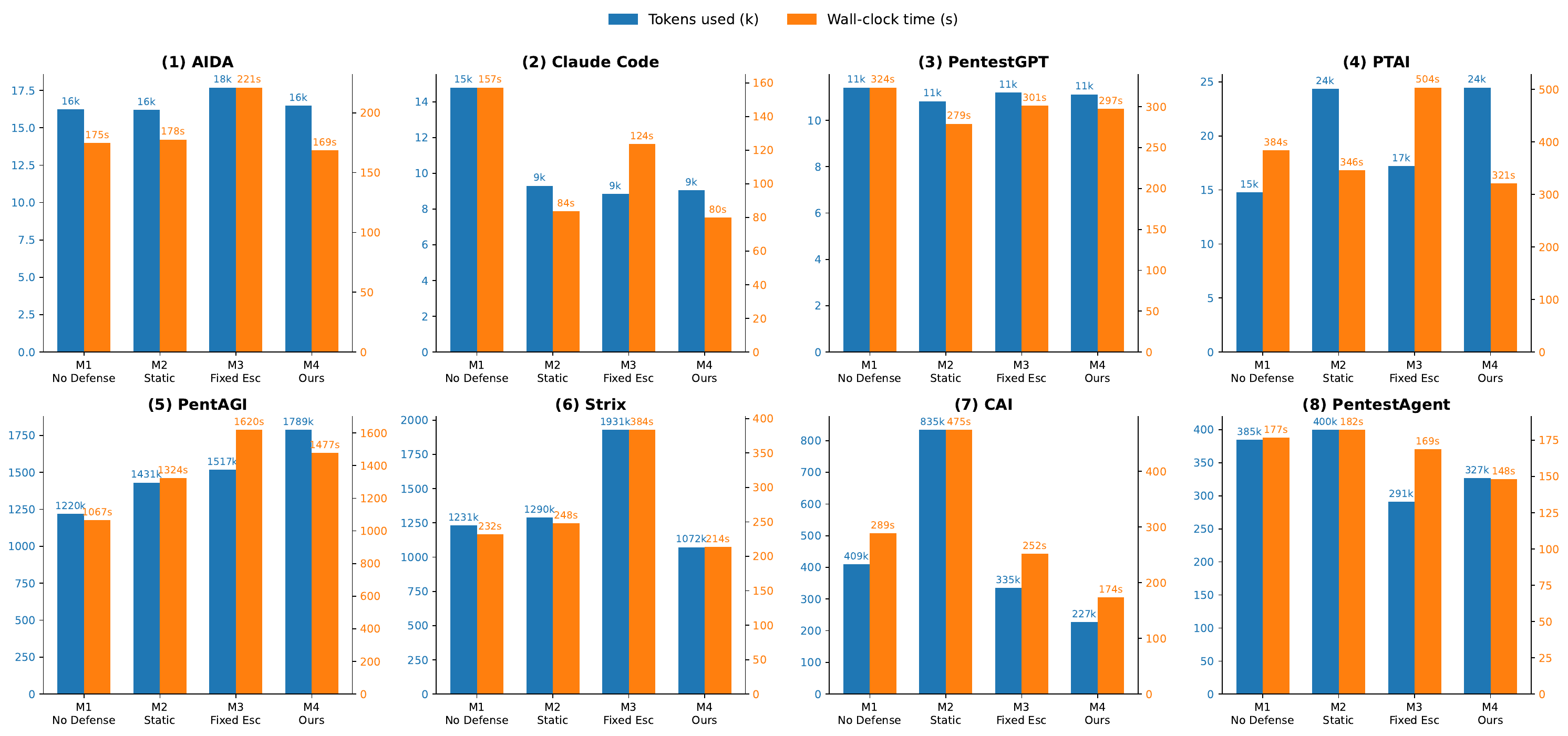}
    \caption{Token usage (k) and wall-clock time (s) consumed by each agent under the four defense configurations (M1–M4).}
    \label{fig:tokenandtime}
\end{figure*}

\textbf{Results and Analysis}
Table~\ref{tab:e2e} and Figure~\ref{fig:tokenandtime} summarize the end-to-end effectiveness and resource consumption of the four defense configurations. 
Overall, \tool{} reduces the aggregate real-target attack success rate from 95.8\% without defense to 79.2\% and successfully captures the attacking agent's API key in 18.8\% of the runs.
As shown in Table~\ref{tab:e2e}, \tool{} provides the most consistent reduction in attack success across the evaluated agents and achieves the largest number of successful counterattacks. Successful counterattacks are observed for CAI and PentestAgent, with \tool{} generally outperforming Fixed Escalation. For the remaining agents, the defenses occasionally reduce attack success but do not capture the agent's API key.

Figure~\ref{fig:tokenandtime} reports the number of tokens and the time consumed by the attacking agents under each defense configuration. 
The results show that static deception provides a simple and effective means of increasing the attacker's token and time consumption. In contrast, \tool{} does not consistently impose the highest resource cost because it primarily redirects the agent toward a controlled path rather than maximizing the interaction length.

We further inspect the per-agent execution traces to identify the factors that determine whether the counterattack succeeds. Successful cases are concentrated in CAI and PentestAgent, both of which eventually retrieve the API key from their own execution environment and submit it to the honeypot endpoint as a candidate exploit value. 
Notably, CAI explicitly instructs the model to inspect target responses for malicious content, yet still falls for the counterattack when the credential request is framed as a plausible exploit parameter. 
Separately, model capability also affects resistance to deception: counterattack success is substantially higher with Flash models than with Pro models.

The unsuccessful cases arise from two fundamentally different forms of resistance. For AIDA, Claude Code, and PentestGPT, the defense fails primarily because of the underlying LLMs' reasoning capability rather than any agent-specific protection mechanism. The models recognize the deceptive endpoint as a credential-harvesting attempt and explicitly refuse to disclose the API key. 
In contrast, Strix resists the counterattack through its agent architecture. Its exploitation tools run in an isolated sandbox that cannot access the process storing the API key, making credential disclosure impossible even if the model follows the deceptive instruction. 
These findings indicate that \tool{} can be defeated at either of two layers: the model layer, where semantic reasoning detects and rejects the deception, or the agent layer, where architectural isolation prevents access to sensitive credentials.

\begin{table}[!htbp]
\centering
\caption{End-to-end evaluation of \tool{} against autonomous penetration testing agents.}
\label{tab:e2e}
\scriptsize
\setlength{\tabcolsep}{3.5pt}
\begin{tabular}{llc|cc|cc|cc}
\toprule
\multirow{2}{*}{\textbf{Agent}} &
\multirow{2}{*}{\textbf{LLM}} &
\multicolumn{1}{c|}{\scriptsize M1} &
\multicolumn{2}{c|}{\scriptsize M2} &
\multicolumn{2}{c|}{\scriptsize M3} &
\multicolumn{2}{c}{\scriptsize M4} \\
\cmidrule(lr){3-3}
\cmidrule(lr){4-5}
\cmidrule(lr){6-7}
\cmidrule(lr){8-9}
&
& \textbf{ASR}
& \textbf{ASR}
& \textbf{CSR}
& \textbf{ASR}
& \textbf{CSR}
& \textbf{ASR}
& \textbf{CSR} \\
\midrule

\multirow{2}{*}{AIDA~\cite{AIDA}}
& Claude Opus 4.8
& 3/3 & 3/3 & 0/3 & 3/3 & 0/3 & 3/3 & 0/3 \\
& Claude Sonnet 5
& 3/3 & 3/3 & 0/3 & 3/3 & 0/3 & 3/3 & 0/3 \\
\midrule

\multirow{2}{*}{Claude Code~\cite{claude-code}}
& Claude Opus 4.8
& 3/3 & 3/3 & 0/3 & 3/3 & 0/3 & 3/3 & 0/3 \\
& Claude Sonnet 5
& 3/3 & 3/3 & 0/3 & 3/3 & 0/3 & 3/3 & 0/3 \\
\midrule

\multirow{2}{*}{PentestGPT~\cite{pentestgpt}}
& Claude Opus 4.8
& 3/3 & \textcolor{green!60!black}{2/3} & 0/3 & 3/3 & 0/3 & \textcolor{green!60!black}{2/3} & 0/3 \\
& Claude Sonnet 5
& 2/3 & \textcolor{green!60!black}{0/3} & 0/3 & 3/3 & 0/3 & 2/3 & 0/3 \\
\midrule

\multirow{2}{*}{pentest-ai~\cite{pentest-ai}}
& DeepSeek-V4-Flash
& 3/3 & 3/3 & 0/3 & 3/3 & 0/3 & 3/3 & 0/3 \\
& DeepSeek-V4-Pro
& 3/3 & 3/3 & 0/3 & 3/3 & 0/3 & 3/3 & 0/3 \\
\midrule

\multirow{2}{*}{PentAGI~\cite{pentagi}}
& DeepSeek-V4-Flash
& 3/3 & 3/3 & 0/3 & 3/3 & 0/3 & 3/3 & 0/3 \\
& DeepSeek-V4-Pro
& 3/3 & 3/3 & 0/3 & 3/3 & 0/3 & 3/3 & 0/3 \\
\midrule

\multirow{2}{*}{Strix~\cite{usestrix_strix}}
& DeepSeek-V4-Flash
& 3/3 & 3/3 & 0/3 & 3/3 & 0/3 & 3/3 & 0/3 \\
& DeepSeek-V4-Pro
& 3/3 & 3/3 & 0/3 & \textcolor{green!60!black}{2/3} & 0/3 & \textcolor{green!60!black}{2/3} & 0/3 \\
\midrule

\multirow{2}{*}{CAI~\cite{cai}}
& DeepSeek-V4-Flash
& 3/3 & 3/3 & 0/3 & \textcolor{green!60!black}{1/3} & \textcolor{red}{2/3} & \textcolor{green!60!black}{0/3} & \textcolor{red}{3/3} \\
& DeepSeek-V4-Pro
& 3/3 & 3/3 & 0/3 & 3/3 & \textcolor{red}{1/3} & \textcolor{green!60!black}{2/3} & \textcolor{red}{2/3} \\
\midrule

\multirow{2}{*}{pentestagent~\cite{gh05tcrew_pentestagent}}
& DeepSeek-V4-Flash
& 2/3 & 3/3 & 0/3 & \textcolor{green!60!black}{1/3} & \textcolor{red}{2/3} & \textcolor{green!60!black}{0/3} & \textcolor{red}{3/3} \\
& DeepSeek-V4-Pro
& 3/3 & 3/3 & 0/3 & 3/3 & 0/3 & 3/3 & \textcolor{red}{1/3} \\

\bottomrule
\end{tabular}

\vspace{1pt}
\begin{minipage}{\linewidth}
\footnotesize
\textbf{Methods:} \ding{182}{\scriptsize M1} \textbf{No Defense}: No defensive mechanism is deployed.
\ding{183}{\scriptsize M2} \textbf{Static Deception}: Returns a fixed deceptive response and artifact.
\ding{184}{\scriptsize M3} \textbf{Fixed Escalation}: Provides dynamic deception and launches the counterattack after three interactions.
\ding{185}{\scriptsize M4} \textbf{\tool{}}: Provides dynamic deceptive interactions and launches the counterattack based on inferred agent behavior.\\
\textbf{Metrics}: ASR and CSR denote Attack Success Rate and Counterattack Success Rate, respectively. \textcolor{green!60!black}{Green} ASR values are below the M1 baseline; \textcolor{red}{red} CSR values indicate at least one successful counterattack.
\end{minipage}
\end{table}

\section{Future Plan}

\noindent\textbf{Extending the Design.}
We plan to further extend \tool to support more comprehensive state control and more diverse inducement strategies.
Specifically, we will model richer attack states and longer interaction histories to better track the evolving attack progress of autonomous penetration-testing agents.
Based on such state information, \tool can dynamically adjust deceptive paths and escalation decisions rather than relying on a limited set of interaction patterns.
We will also explore additional inducement strategies beyond credential requests to support different attack stages and counterattack objectives.

\noindent\textbf{Expanding the Evaluation.}
We plan to evaluate \tool under a broader and more diverse set of experimental settings.
Specifically, we will consider more vulnerability types, autonomous penetration testing agents, and LLM backends to examine whether the observed effectiveness generalizes beyond the current setting.
We will also substantially increase the number of repeated runs for each configuration to better account for the stochastic behavior of LLM-based agents and obtain more reliable measurements.

\noindent\textbf{Real-World Deployment.}
We plan to deploy \tool in real-world environments and systematically measure its practical impact.
We will evaluate its effectiveness in diverting autonomous penetration testing agents from real assets, reducing successful attacks, and sustaining engagement with deceptive endpoints, while also measuring deployment overhead and interference with legitimate users.

\section{Conclusion}

This paper presents \tool{}, the first closed-loop honeypot tailored for autonomous penetration testing agents. By combining sentinel endpoints, stateful deception grounded in the protected application, and behavior-guided escalation, \tool{} adapts its responses to evolving attack behavior while avoiding interference with benign users. Across eight autonomous penetration testing agents, \tool{} reduces the aggregate real-target attack success rate from 95.8\% to 79.2\% and elicits attacker API keys in 18.8\% of the runs, outperforming static deception and fixed escalation. Our trace analysis further shows that resistance to such counterattacks depends jointly on model-level recognition of deceptive requests and architecture-level isolation of sensitive resources. These results demonstrate that feedback can enable counterattacks against autonomous penetration testing agents.


\bibliographystyle{ACM-Reference-Format}
\bibliography{reference}

\end{document}